\documentclass{article}
\usepackage{spconfa4,amsmath,graphicx}

\title{Weakly Guided and Autoregressive Beamformer Parameterization for Generalizable Moving Speaker Extraction in Higher-Order Ambisonics}

\name{Jakob Kienegger, Tal Peer, Sina Khanagha, Timo Gerkmann\thanks{
This work was supported by the Deutsche Forschungsgemeinschaft (DFG) under Grant 508337379. Computational resources were provided by the Regional Computer Center (RRZ) of the University of Hamburg and the Erlangen National High Performance Computing Center (NHR@FAU) under Project f104ac. NHR is funded by the Federal Government and Bavaria. Both facilities received DFG funding under Grants 440719683 and 498394658.
}}

\address{
    Signal Processing (SP) Group, University of Hamburg
}

\usepackage{comment}
\usepackage[nolist, nohyperlinks]{acronym}
\usepackage{nicefrac}
\usepackage{booktabs}
\usepackage{pifont}
\usepackage{cite}

\usepackage{amsmath,graphicx}
\usepackage[hidelinks]{hyperref}
\usepackage[capitalize,nameinlink]{cleveref}
\usepackage{amssymb}
\usepackage{dsfont}
\usepackage{circuitikz}
\usepackage[dvipsnames]{xcolor}
\usepackage{tikz}
\usetikzlibrary{decorations.pathreplacing,calligraphy, arrows.meta, shapes.geometric, positioning, shapes.misc, calc}
\usepackage{pgf}
\usepackage{pgfplots}
\usepackage{contour}
\contourlength{0.4pt}
\tikzset{cross/.style={cross out, draw=black, minimum size=2*(#1-\pgflinewidth), inner sep=0pt, outer sep=0pt},}
\usepackage{subcaption}
\crefformat{equation}{(#2#1#3)}
\Crefformat{equation}{(#2#1#3)}
\crefmultiformat{equation}
  {(#2#1#3)} %
  {, (#2#1#3)} %
  {, (#2#1#3)} %
  {, (#2#1#3)} %

\Crefmultiformat{equation}
  {(#2#1#3)}
  {, (#2#1#3)}
  {, (#2#1#3)}
  {, (#2#1#3)}
\usepackage{siunitx}
\pgfplotsset{compat=1.18}

\DeclareMathAlphabet\mathbfcal{OMS}{cmsy}{b}{n}

\newcommand\projectPage{
    \scriptsize \texttt{https://sp-uhh.github.io/weakly-guided-beamforming/}
}

\acrodefplural{nns}[NNs]{neural networks}
\acrodefplural{irm}[IRMs]{ideal ratio masks}
\acrodefplural{ibm}[IBMs]{ideal binary masks}
\acrodefplural{doa}[DoAs]{directions of arrival}
\acrodefplural{rirs}[RIRs]{room impulse responses}

\begin{acronym}
\acro{stft}[STFT]{short-time Fourier transform}
\acro{rir}[RIR]{room impulse response}
\acro{istft}[iSTFT]{inverse short-time Fourier transform}
\acro{doa}[DoA]{direction of arrival}
\acro{irm}[IRM]{ideal ratio mask}
\acro{mvdr}[MVDR]{minimum-variance distortionless response}
\acro{nn}[NN]{neural network}
\acro{dnn}[DNN]{deep neural network}
\acro{gcc_phat}[GCC-PHAT]{generalized cross-correlation phase transform}
\acro{ssl}[SSL]{sound source localization}
\acro{srp_phat}[SRP-PHAT]{steered response power with phase transform}
\acro{mse}[MSE]{mean squared error}
\acro{mae}[MAE]{mean angular error}
\acro{sdr}[SDR]{signal-to-distortion ratio}
\acro{sisdr}[SI-SDR]{scale-invariant signal-to-distortion ratio}
\acro{pesq}[PESQ]{perceptual evaluation of speech quality}
\acro{polqa}[POLQA]{perceptual objective listening quality}
\acro{wer}[WER]{word error rate}
\acro{pit}[PIT]{permutation-invariant training}
\acro{lbt}[LBT]{location-based training}
\acro{asr}[ASR]{automatic speech recognition}
\acro{bce}[BCE]{binary cross-entropy}
\acro{jnf}[JNF]{joint non-linear filter}
\acro{nn}[NN]{neural network}
\acro{dnn}[DNN]{deep neural network}
\acro{mccruse}[MC-CRUSE]{multi-channel convolutional recurrent U-net architecture for speech enhancement}
\acro{estoi}[ESTOI]{extended short-time objective intelligibility}
\acro{snr}[SNR]{signal-to-noise ratio}
\acro{sir}[SIR]{signal-to-interference ratio}
\acro{iid}[i.i.d.]{independent and identically distributed}
\acro{wrt}[w.r.t.]{with respect to}
\acro{ssf}[SSF]{spatially selective filter}
\acro{cv}[CV]{constant velocity}
\acro{rw}[RW]{random walk}
\acro{ar}[AR]{autoregressive}
\acro{tse}[TSE]{target speaker extraction}
\acro{tst}[TST]{target speaker tracking}
\acro{tsl}[TSL]{target speaker localization}
\acro{dnsmos}[DNSMOS]{deep noise suppression mean opinion score}
\acro{map}[MAP]{maximum a posteriori}
\acro{kf}[KF]{Kalman filter}
\acro{pf}[PF]{particle filter}
\acro{das}[DaS]{delay-and-sum}
\acro{ae}[AE]{angular error}
\acro{acc}[ACC]{accuracy}
\acro{miso}[MISO]{multiple-input and single-output}
\acro{mimo}[MIMO]{multiple-input and multiple-output}
\acro{foa}[FOA]{first order ambisonics}
\acro{hoa}[HOA]{higher-order ambisonics}
\acro{sh}[SH]{spherical harmonics}
\acro{sht}[SHT]{spherical harmonics transform}
\acro{mwf}[MWF]{multichannel wiener filter}
\acro{seld}[SELD]{sound event localization and detection}
\acro{mac}[MAC]{multiply-accumulate operation}
\acro{acn}[ACN]{ambisonics channel number ordering}
\acro{rds}[RDS]{recurrent deep stacking}
\acro{mos}[MOS]{mean opinion score}
\acro{nisqa}[NISQA]{non-intrusive speech quality assessment}
\acro{pev}[PEV]{principal eigenvector}
\acro{fb}[FB]{fixed beamformer}
\acro{ema}[EMA]{exponential moving average}
\end{acronym}

\definecolor{tab_blue}{RGB}{31, 119, 180}
\definecolor{tab_green}{RGB}{44, 160, 44}
\definecolor{tab_orange}{RGB}{255, 127, 14}
\definecolor{tab_purple}{RGB}{148, 103, 189}
\definecolor{tab_brown}{RGB}{165, 42, 38}
\definecolor{tab_pink}{RGB}{227, 119, 194}
\definecolor{tab_cyan}{RGB}{23, 190, 207}
\definecolor{orig_green}{RGB}{44, 160, 44}
\definecolor{orig_purple}{RGB}{148, 103, 189}
\definecolor{custom_green}{RGB}{149.5, 207.5, 149.5}
\definecolor{custom_purple}{RGB}{201.5, 179, 222}
\definecolor{dark_gray}{RGB}{128, 128, 128}

\newcommand\majorTick{0.75mm}
\newcommand\minorTick{0.5mm}
\newcommand\tickFont{\footnotesize}

\newcommand\myPhi{\mathbf{\Phi}}
\newcommand\speechNoise{\xi}
\newcommand\speechNoiseBold{\boldsymbol{\xi}}

\newcommand\tickWidth{0.5pt}

\newcommand\elevation{\phi}

\newcommand\delayIdx[1]{
    \scriptstyle #1\hspace{0.25mm}\raisebox{0.4ex}{\rule{0.3em}{0.06ex}}1
}
\newcommand\smallDelayIdx[1]{
    \scriptstyle #1\hspace{0.25mm}\raisebox{0.4ex}{\rule{0.4em}{0.2ex}}1
}
\newcommand\myPM{\raisebox{0.3ex}{\tiny\ensuremath{\pm}}}

\newcommand\transposed{{\mathsf T}}
\newcommand\hermitian{{\mathsf H}}
\newcommand\inverse{{{\hspace{-0.5mm}\scalebox{0.5}[1.0]{$-$}\hspace{-0.0mm}\scalebox{0.7}{$1$}}}}
\newcommand\hoaFirstCoeff{{{1\hspace{0.25mm}\scalebox{0.5}[1.0]{$-$}\hspace{-0.0mm}\scalebox{0.7}{$1$}}}}
\newcommand\superscript[1]{{\scriptscriptstyle{#1}}}

\newcommand\splitSize{3pt}
\newcommand\addSize{6pt}
\newcommand\shadowOff{0.375mm}
\newcommand\smallMath{\tiny}
\newcommand\myDelta{\Delta\hspace{-0.25mm}}
\newcommand\ambixWidth{0.5pt}
\newcommand\rotEllipseA{0.27cm} %
\newcommand\rotEllipseB{0.12cm}  %
\newcommand{\rotSymb}[1]{
     \makebox[1cm][l]{\hspace{-1.15mm}\raisebox{0mm}{
        \begin{tikzpicture}
            \draw[x=\rotEllipseB, y=0.95*\rotEllipseA, line width=0.175ex, -stealth, color=lightgray] ({1*cos(-150)},{1*sin(-150)}) arc (-150:150:1 and 1);
            \draw[x=\rotEllipseA, y=0.7*\rotEllipseB, line width=0.175ex, -stealth, color=lightgray] ({1*cos(-150)},{1*sin(-150)}) arc (-150:150:1 and 1);
            \node[anchor=center] at (0,0) {\scriptsize\contour{white}{\hspace{0.5mm}#1}};
        \end{tikzpicture}
    }
    }
}
\tikzstyle{rotBox} = [
  rectangle,
  draw=black,
  minimum width=0.6cm,
  minimum height=0.6cm,
  inner sep=0pt,
  outer sep=0pt,
  align=center,
  text width=0.6cm,
  text height=0.6cm,
  text depth=0cm
]

\def\mvdrW{0.6cm}
\def\mvdrH{0.4cm}
\tikzstyle{mvdrBox} = [
  rectangle,
  draw=black,
  minimum width=\mvdrW,
  minimum height=\mvdrH,
  inner sep=0pt,
  outer sep=0pt,
  align=center,
  text width=\mvdrW,
  text height=\mvdrH,
  text depth=0cm
]
\def\mvdrSymbH{0.35cm}
\def\mvdrSymbW{0.6cm}
\newcommand\mvdrSymb{
    \includegraphics[height=\mvdrSymbH, width=\mvdrSymbW]{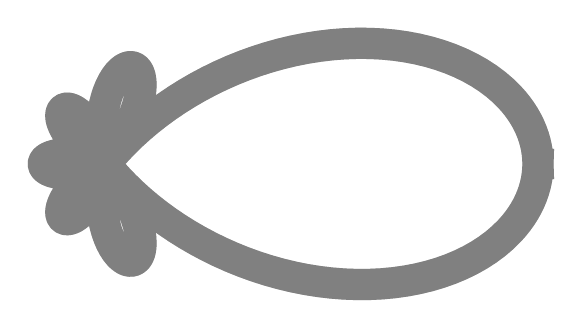}
}

\def\normW{0.45cm} %
\def\normH{0.3cm}
\tikzstyle{normBox} = [
  rectangle,
  draw=black,
  minimum width=\normW,
  minimum height=\normH,
  inner sep=0pt,
  outer sep=0pt,
  align=center,
  text width=\normW,
  text height=\normH,
  text depth=0cm,
  fill=white
]
\newcommand{\normSymb}{
  \makebox[1cm][l]{\hspace{-1mm}\raisebox{0.25mm}{
  \begin{tikzpicture}
  \node {\tiny$\lVert\hspace{-0.1mm}\cdot\hspace{-0.1mm}\rVert^{\hspace{-0.2mm}\scalebox{0.8}{$2$}}$};
  \end{tikzpicture}
  }
  }
}

\tikzstyle{emaBox} = [
  rectangle,
  draw=black,
  minimum width=0.95cm, %
  minimum height=0.65cm,
  inner sep=0pt,
  outer sep=0pt,
  align=center,
  text width=0.6cm,
  text height=0.4cm,
  text depth=0cm
]
\def\circleR{1.5} 
\tikzset{
  multi/.style={
    draw,
    circle,
    minimum size=2*\circleR,
    inner sep=0pt,             %
    outer sep=0pt,             %
    path picture={
      \draw
        ($(path picture bounding box.center) + (-0.77*\circleR, -0.77*\circleR)$)
        --
        ($(path picture bounding box.center) + ( 0.77*\circleR,  0.77*\circleR)$);
      \draw
        ($(path picture bounding box.center) + ( 0.77*\circleR, -0.77*\circleR)$)
        --
        ($(path picture bounding box.center) + (-0.77*\circleR,  0.77*\circleR)$);
    }
  }
}
\tikzset{
  adder/.style={
    draw,
    circle,
    minimum size=2*\circleR,
    inner sep=0pt,
    path picture={
      \draw
        ($(path picture bounding box.center) + (-0.35*\circleR, 0)$)
        --
        ($(path picture bounding box.center) + ( 0.35*\circleR, 0)$);
      \draw
        ($(path picture bounding box.center) + (0, -0.35*\circleR)$)
        --
        ($(path picture bounding box.center) + (0,  0.35*\circleR)$);
    }
  }
}
\tikzset{
  basenode/.style={
    minimum size=2*\circleR,
    inner sep=0pt
  }
}
\tikzstyle{smallSplit} = [circle, fill=gray, minimum size=\circleR, inner sep=0pt]
\def\arrowsize{1mm}
\tikzset{
  smallarrow/.style={
    -{Latex[length=\arrowsize,width=\arrowsize]},
    line width=0.4pt
  }
}
\newcommand\nodeLen{3mm}
\newcommand{\emaSymb}[1]{%
\makebox[1cm][l]{\hspace{-2.5mm}%
\raisebox{-0.5mm}{%
\begin{tikzpicture}[scale=0.55, draw=gray, text=gray]
    
    \node[basenode] (origin) {};

    \node[basenode] (noisy_in) at (origin) {};
    \node[basenode] (mask_in) [above=\nodeLen/2 of noisy_in] {};
    \node[multi, right=\nodeLen/2 of noisy_in] (multi1) {};
    \node[adder, right=\nodeLen of multi1] (add) {};
    \node (multi2) [multi, above=\nodeLen/2 of add] {};
    \node[basenode, xshift=-4.5mm, yshift=1mm] at (multi2.center) {\tiny$1\hspace{-0.15mm}\raisebox{0.3em}{\rule{0.4em}{0.2ex}}\gamma$};
    \node[smallSplit, xshift=-\circleR/2] (mask_split) at (multi1.center |- mask_in.center) {};
    \node[basenode, xshift=-2mm, yshift=-1.25mm] at (multi1.center |- mask_split.center) {\tiny$\gamma$};
    \node[basenode] (cov_in) [above=\nodeLen/2 of multi2] {};
    \node[basenode] (cov_out) [right=\nodeLen/2 of add] {};

    \draw[smallarrow] (noisy_in) -- (multi1);
    \draw[smallarrow] (multi1) -- (add);
    \draw[smallarrow] (multi2) -- (add);
    \draw[smallarrow] (mask_in) -- (multi2);
    \draw[smallarrow] (mask_split) -- (multi1);
    \draw[smallarrow] (cov_in) -- (multi2);
    \draw[smallarrow] (add) -- (cov_out);
    
\end{tikzpicture}%
}}}
\newcommand{\dnnSymb}[1]{%
  \makebox[1.5cm][l]{\hspace{-1.05mm}
  \begin{tikzpicture}

    \def\r{0.25mm} %
    \def\xleft{-0.3}
    \def\xmidl{-0.125}
    \def\xmidr{0.125}
    \def\xright{0.3}
    \def\yspace{1.9mm} %
    \def\numNmid{7}
    \def\numNin{5}
    
    \foreach \i in {1,..., \numNin} {
      \node[anchor=center, circle,fill=gray,inner sep=0pt,minimum size=2*\r] (ll\i) at (\xleft,{\yspace*(\i-\numNin+ (\numNmid - \numNin) / 2)}) {};
    }
    \foreach \i in {1,...,\numNmid} {
      \node[anchor=center, circle,fill=gray,inner sep=0pt,minimum size=2*\r] (ml\i) at (\xmidl,{\yspace*(\i-\numNin)}) {};
    }
    \foreach \i in {1,...,\numNmid} {
      \node[anchor=center, circle,fill=gray,inner sep=0pt,minimum size=2*\r] (mr\i) at (\xmidr,{\yspace*(\i-\numNin)}) {};
    }
    \foreach \i in {1,..., \numNin} {
      \node[anchor=center, circle,fill=gray,inner sep=0pt,minimum size=2*\r] (r\i) at (\xright,{\yspace*(\i-\numNin+ (\numNmid - \numNin) / 2)}) {};
    }
    \foreach \li in {1,..., \numNin}{
      \foreach \mi in {1,...,\numNmid}{
        \draw[anchor=center, lightgray,line width=0.25pt] (ll\li) -- (ml\mi);}
    }

    \foreach \li in {1,..., \numNmid}{
      \foreach \mi in {1,...,\numNmid}{
        \draw[anchor=center, lightgray,line width=0.25pt] (ml\li) -- (mr\mi);}
    }
    \foreach \mi in {1,...,\numNmid}{
      \foreach \ri in {1,...,\numNin}{
        \draw[anchor=center, lightgray,line width=0.25pt] (mr\mi) -- (r\ri);}
    }
    \node[anchor=center] at (0, 0) {\contour{white}{#1}};
  \end{tikzpicture}
  }
}
\def\dnnW{0.8cm}
\def\dnnH{1.25cm}
\tikzstyle{dnnBox} = [
  rectangle,
  draw=black,
  minimum width=\dnnW,
  minimum height=\dnnH,
  inner sep=0pt,
  outer sep=0pt,
  align=center,
  text width=\dnnW,
  text height=\dnnH,
  text depth=0cm
]

\tikzstyle{delay} = [
  rectangle,
  draw=black,
  minimum width=0.3cm,
  minimum height=0.3cm,
  inner sep=0pt,
  outer sep=0pt,
  align=center,
  text width=0.3cm,
  text height=0.3cm,
  text depth=0cm,
  fill=white
]
\newcommand{\delaySymb}{
  \makebox[1cm][l]{\hspace{-1mm}\raisebox{0.25mm}{
  \begin{tikzpicture}
  \node {\tiny$z^{\hspace{-0.5mm}\scalebox{0.5}[1.0]{$-$}\hspace{-0.25mm}\scalebox{0.8}{$1$}}$};
  \end{tikzpicture}
  }
  }
}
\tikzstyle{split} = [circle, fill=black, minimum size=\splitSize, inner sep=0pt]
\tikzstyle{add} = [circle, fill=white, minimum size=\addSize, inner sep=0pt, draw=black]
\tikzstyle{arrow} = [line width=\ambixWidth,
    -{Computer Modern Rightarrow[line cap=round,width=6pt,length=2.5pt,line width=1pt]},
    draw=black]

\newcommand{\picLegend}[1]{\raisebox{-0.275ex}{\includegraphics[height=0.75em]{#1}}}  %
\newcommand{\picTable}[1]{\raisebox{-0.275ex}{\includegraphics[height=0.9em]{#1}}}  %

\usepackage{eso-pic}

\begin{document}
\ninept
\maketitle
\begin{abstract}
Linear spatial filters (beamformers) enable robust, generalizable and interpretable speech enhancement with performance guarantees under ideal parameterization.
Modern beamformers are often parameterized by deep neural networks, whose performance degrades in dynamic scenarios with multiple moving speakers of unknown directions.
We propose a data-driven beamforming pipeline, which only requires an estimate of the target’s initial direction.
Building on a higher-order ambisonics representation, we show that neural temporal-spectral processing can be decoupled from linear spatial processing, and thereby achieve generalizable and array-agnostic enhancement.
By incorporating autoregression into a frame-wise causal framework, we maintain consistent performance throughout fast speaker motion and long recordings.
Evaluation on synthetic data demonstrates robust enhancement under challenging conditions with closely spaced and crossing speakers.
Real-world recordings in a dynamic office meeting scenario complement these findings and show generalizability across varying ambisonics orders.

\end{abstract}


\section{Introduction}
\label{sec:intro}
Speech enhancement aims to improve the perceptual quality and intelligibility of a recorded speech signal by suppressing noise and reverberation. 
In a multi-speaker environment, \ac{tse} solves the additional challenge of differentiating between the desired target speech and remaining speakers, which are to be treated as noise.
When a recording from a microphone array is available, the relative direction of the target to the array, known as \ac{doa}, can be employed to identify the desired speaker.
In case of a spherical array, ambisonics~\cite{zotter19ambisonics} provides a \textit{directional} and \textit{array-agnostic} representation of the recorded sound field, making it a popular audio format for enhancement~\cite{perotin18lstm_ambisonics_beamforming,herzog20direction_reverberation_noise_reduction,lugasi20ambisonics_masking_binaural_reproduction,kienegger25deep_joint_ar_ssf}.

Assuming the target's \ac{doa} is known, a spatial filter can be steered toward the desired direction and extract the corresponding speech signal.
By jointly processing temporal-spectral and spatial information, recently proposed deep, non-linear spatial filters achieve outstanding enhancement performance \cite{tesch24ssf_journal,bohlender24sep_journal}.
While linear spatial filters, known as beamformers, are proven to be inferior under realistic, non-Gaussian statistical assumptions~\cite{tesch21nonlinear_spatial_filtering},
they provide \textit{robustness} and \textit{interpretability} by being a parametric, statistics-based approach.
When combined with \acp{dnn} to estimate these statistics, such hybrid approaches can achieve strong enhancement performance even with linear spatial filters~\cite{heymann16neural_beamformer,umbach24hybrid_approaches_parameter_estimation_filtering}.

In stationary scenarios, the target's \ac{doa} may be available a priori. 
However, when speaker motion becomes non-negligible, the continuous directional information necessary to steer a spatial filter---which we refer to as \textit{strong} guidance---is typically not available.
\textit{Weakly} guided speaker extraction relaxes this constraint and only assumes knowledge of the target's direction at recording start.
To continue using a spatial filter for extraction, an additional tracking algorithm is typically required to infer the target's movement from the starting direction and automate the steering of the spatial filter throughout the remaining recording~\cite{kienegger25wg_ssf}.
Nevertheless, to maintain precise target alignment in challenging acoustic scenarios with closely spaced and crossing speakers, computationally heavy, data-driven tracking algorithms become necessary~\cite{bohlender21ssl_temporal_context}.

Besides explicit directional steering, estimation of the spatial statistics required for beamformer parameterization can also be achieved implicitly via \ac{dnn}-driven signal indicator masks.
These masks solely contain temporal-spectral information and are used to improve the performance of classical statistical moment estimation techniques.
Without explicitly modeling \ac{doa} trajectories, beamformer parameterization becomes unaffected by closely spaced speakers and can even operate without learning the microphone array geometry~\cite{boeddeker24ts_sep}.
This results in 
generalizability across diverse motion profiles~\cite{ochiai23moving_speaker_attention_mvdr} as well as acoustic and recording conditions.

Streaming real-time applications 
require causal frame-wise processing, which limits the input to current and past frames. However, the availability of past frames allows for exploitation of \textit{temporal-spectral} correlations, resulting in  improved enhancement performance~\cite{pan24paris_autoregressive_separation,shen25arise}.
In our prior work, we demonstrated how leveraging previous frames in an \ac{ar} framework based on a deep, non-linear spatial filter can significantly increase speaker separability during spatially challenging acoustic scenarios~\cite{kienegger25deep_joint_ar_ssf}. 

In this work, we focus on linear spatial filtering, i.e. beamforming, for weakly guided moving speaker extraction in \ac{hoa} recordings.
We employ a \ac{dnn}-driven mask to estimate the parameters of a beamformer and decouple mask computation from spatial processing to remain ambisonics-order-agnostic.
Motivated by prior work with stationary speakers~\cite{nugraha16source_separation_das_input,perotin18lstm_ambisonics_beamforming,elminshawi23bf_guided_tse}, we convert the target's initial direction into an input feature for the \ac{dnn}. 
Additionally, based on an underlying frame-wise causal processing pipeline, we employ temporal feedback of the enhanced speech signal to increase robustness and continuity of the estimated masks while retaining streaming capability.
Evaluation on synthetic noisy and reverberant two-speaker mixtures demonstrates how our \ac{ar} framework robustly extracts the moving target while maintaining consistent performance throughout
long audio recordings, i.e. exceeding 30\,s. 
Processed real-world recordings confirm the generalization capability of our methods to 
challenging acoustic conditions in an office meeting scenario across varying ambisonics orders.

\section{Problem Setup}
\label{sec:problem}
\subsection{Ambisonics signal representation}
Let $\mathcal{X}(r, \theta, \phi)$ denote an acoustic pressure field in spherical coordinates at radius $r$, azimuth $\theta$ and elevation $\phi$ centered about a spherical microphone array.
At microphone position $p$, the recorded sound field can be approximated by the truncated \ac{sh} expansion in the \ac{stft} domain as
\begin{equation}\label{eq:ambisonics_expansion}
    \mathcal{X}_{tk}(r_p, \theta_p, \phi_p) \approx \sum_{n=0}^N \sum_{m=-n}^n b_n(\kappa_k r_p) \, X_{tk}^{nm} \, Y^m_n(\theta_p, \phi_p) \, ,
\end{equation}
with frequency bin $k$ and frame index $t$.
We use the real-valued \acp{sh}
\begin{equation}\label{eq:real_sh}
Y^m_n(\theta, \elevation) = \begin{cases}
    N_n^{\left|m\right|} P_n^{\left|m\right|}\!\left(\sin(\elevation)\right) \cos(m\,\theta) \,, &\text{if} \ \ m \ge 0\,, \\
    N_n^{\left|m\right|} P_n^{\left|m\right|}\!\left(\sin(\elevation)\right) \sin(|m|\theta) \,, &\text{else},
\end{cases}
\end{equation}
with associated Legendre functions $P_n^{m}$ and normalization factor $N_n^m$ (N3D).
The radial response $b_n(\kappa_k r)$ with frequency dependent wavenumber $\kappa_k$ describes the scattering effects of the spherical array. 
At truncation order $N$, the series contains $M=(N+1)^2$ coefficients $X_{tk}^{nm}$, which we denote in the vectorized form $\mathbf{X}_{tk} = \big(X_{tk}^{00}, X_{tk}^\hoaFirstCoeff, \dots, X_{tk}^{N\hspace{-0.25mm}N}\big)^\transposed$. 
These coefficients represent the sound field in an \textit{array-agnostic} audio format, known as ambisonics~\cite{zotter19ambisonics}. 
If $P \geq M$ spatial samples, i.e. microphones, are available, \cref{eq:ambisonics_expansion} can be solved in the least-squares sense to obtain $\mathbf{X}_{tk}$~\cite{moreau06hoa_4th_order,jarrett17theory_applications_sh_processing}.

\subsection{Ambisonics signal model}
Modeling the recorded sound field as a superposition of speech and noise allows decomposing $\mathbf{X}_{tk}$ into ambisonics coefficients corresponding to the target speaker $\mathbf{S}_{tk}$ and noise $\mathbf{V}_{tk}$~[\citen{rafaely19spherical_array_processing}, Sec.~7.1],
\begin{equation}\label{eq:signal_model}
    \mathbf{X}_{tk} = \mathbf{S}_{tk} + \mathbf{V}_{tk} \, \in \mathbb{C}^M \, .
\end{equation}
Assuming a noisy and reverberant multi-speaker scenario, $\mathbf{V}_{tk}$ absorbs interfering speech signals, environmental and measurement noise as well as the reverberant part of the target speech. 
The anechoic target speech signal $\mathbf{S}_{tk}$ can be further factored as $\mathbf{S}_{tk} =\mathbf{d}_t S_{tk}$, with steering vector $\mathbf{d}_t = \mathbf{d}(\theta_t, \phi_t)$ encoding the direct path propagation corresponding to the target's \ac{doa} $\theta_t, \phi_t$.
Under far-field assumptions, $\mathbf{d}_t$ equals the \ac{sh} basis vector $\mathbf{Y}(\theta_t, \phi_t)$ in \cref{eq:real_sh} and is thereby real-valued and frequency-independent~[\citen{rafaely19spherical_array_processing}, Sec.~2.4].

\subsection{Ambisonics beamforming}

For \acl{tse}~(\acs{tse}), we employ a linear spatial filter (beamformer) $\mathbf{w}_{tk}$ to reconstruct the monaural target speech signal $S_{tk}$ from observation $\mathbf{X}_{tk}$.
We parameterize $\mathbf{w}_{tk}$ based on 
second-order statistics of speech and noise signals,
i.e. covariance matrices $\myPhi_{tk}^\superscript{(\mathrm{S})}$ and $\myPhi_{tk}^\superscript{(\mathrm{V})}$, whose estimation is at the center of this work.

In particular, we choose an \ac{mvdr} beamformer, which suppresses noise under a unit gain and phase-preserving  (distortionless) constraint toward the target \ac{doa}.
The \ac{mvdr} beamformer $\mathbf{w}_{tk}$ can be formulated as
\begin{equation}\label{eq:miso_mvdr}
    \mathbf{w}_{tk} = \frac{\big(\myPhi_{tk}^\superscript{(\mathrm{V})}\big)^\inverse\mathbf{d}_t}{\mathbf{d}_t^\transposed\big(\myPhi_{tk}^\superscript{(\mathrm{V})}\big)^\inverse\mathbf{d}_t} \quad \text{with} \quad \hat{S}_{tk} = \mathbf{w}_{tk}^\hermitian \mathbf{X}_{tk} \, .
\end{equation}
For parameterization, a frequency-dependent estimate of the steering vector $\hat{\mathbf{d}}_{tk}$ can be obtained via the \ac{pev} of the speech covariance matrix $\myPhi_{tk}^\superscript{(\mathrm{S})}$.
Since the \ac{pev} is phase-invariant, we follow~\cite{jarrett17theory_applications_sh_processing,herzog20direction_reverberation_noise_reduction} and align $\hat{\mathbf{d}}_{tk}$ at the omnidirectional channel and adjust its scaling such that $4 \pi \lVert \hat{\mathbf{d}}_{tk} \rVert^2 = M$ [\citen{rafaely19spherical_array_processing}, Sec.~1.2].

\section{Proposed Method}
 \label{sec:method}
\subsection{Mask-based recursive covariance matrix estimation}

To infer the time-varying speech and noise covariance matrices $\myPhi_{tk}^\superscript{(\mathrm{S})}$ and $\myPhi_{tk}^\superscript{(\mathrm{V})}$, we employ recursive covariance estimation of the form
\begin{equation}\label{eq:covariance_estimation}
    \hat{\myPhi}_{tk}^\superscript{(\mathrm{\speechNoise})} = \gamma_{tk}^\superscript{(\mathrm{\speechNoise})}\mathbf{X}_{tk}\mathbf{X}_{tk}^\hermitian + \left(1 - \gamma_{tk}^\superscript{(\mathrm{\speechNoise})}\right) \hat{\myPhi}_{\delayIdx{t}k}^\superscript{(\mathrm{\speechNoise})}\, , 
\end{equation}
with $\mathrm{\speechNoise} \in \{\mathrm{S}, \mathrm{V}\}$. 
We define the composite parameters $\gamma_{tk}^\superscript{(\mathrm{\speechNoise})}$ as
\begin{equation}\label{eq:gamma_def}
    \gamma_{tk}^\superscript{(\mathrm{\speechNoise})} = \alpha^\superscript{(\mathrm{\speechNoise})}\hat{\mathcal{M}}_{tk}^\superscript{(\mathrm{\speechNoise})} \Big/ \big(\alpha^\superscript{(\mathrm{\speechNoise})}\hat{\mathcal{M}}_{tk}^\superscript{(\mathrm{\speechNoise})} + 1 - \alpha^\superscript{(\mathrm{\speechNoise})}\big) \, ,
\end{equation}
with estimated soft signal indicator masks $\hat{\mathcal{M}}_{tk}^\superscript{(\mathrm{\speechNoise})}$ and temporal smoothing parameters $\alpha^\superscript{(\mathrm{\speechNoise})}$.
Note that our proposed formulation deviates from conventional mask-based \acl{ema} definitions~\cite{umbach24hybrid_approaches_parameter_estimation_filtering}.
Expanding the recursion in \cref{eq:covariance_estimation,eq:gamma_def} shows how the denominator in \cref{eq:gamma_def} cancels the exponential decay for frames marked without signal activity ($\hat{\mathcal{M}}_{tk}^\superscript{(\mathrm{\speechNoise})}=0$). 
As a result, the effective sample size depends on the mask values $\hat{\mathcal{M}}_{tk}$, which avoids covariance shrinkage in periods of signal inactivity.
For initialization, we use the targets’ starting direction to construct a \mbox{rank-1} speech covariance matrix and a diffuse (\ac{sh}-white~[\citen{rafaely19spherical_array_processing}, Sec.~7.1]) noise model, yielding
\begin{equation}\label{eq:init_covaraince}
    \hat{\myPhi}_{0k}^\superscript{(\mathrm{S})} = \sigma_0^2 \, \mathbf{d}_0\mathbf{d}_0^\transposed \quad \text{and} \quad \hat{\myPhi}_{0k}^\superscript{(\mathrm{V})} = \frac{\sigma_0^2}{4\pi} \, \mathbf{I} \, ,
\end{equation}
with $\sigma_0^2$ adjusted to match the expected input power.
It is to be emphasized that the \textit{adaptive} exponential decay of our proposed recursive covariance estimator in \cref{eq:covariance_estimation,eq:gamma_def} is essential to maintain the target direction throughout periods of speech inactivity at recording start.

\begin{figure}
\makebox[0pt][l]{\hspace*{-25pt}\input{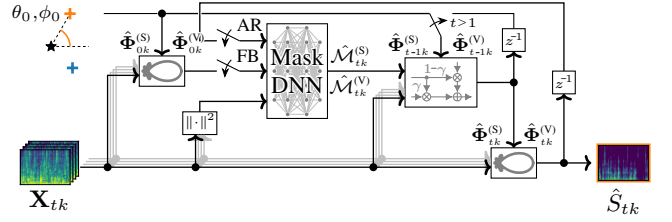}}
    \vspace*{-10pt}
    \caption{Mask-based beamforming for weakly guided speaker extraction using initial \ac{doa} $\theta_0, \phi_0$. A fixed beamformer (FB) and autoregression (AR) is used as conditioning for the mask estimation \ac{dnn}.}
    \label{fig:overview}
    \vspace*{-10pt}
\end{figure}

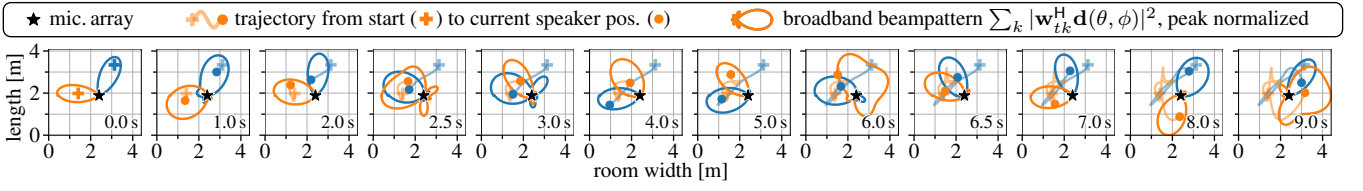
\begin{figure*}[t!]
\begin{tikzpicture}
    
\newcommand\doaWidth{1.5cm}
\newcommand\doaHeight{8mm}
\newcommand\doaDist{2mm}
\newcommand\lrDist{10mm}
\newcommand\tbDist{8mm}
\newcommand\tXoff{3.5mm}
\newcommand\tYoff{-4mm}
\newcommand\plotStartX{0mm}
\newcommand\plotStartY{0mm}
\newcommand\legendTickTop{-0.25mm}
\newcommand\legendTickBottom{-0.5mm}
\pgfmathsetlengthmacro{\legendTickMiddle}{0.5 * \legendTickTop + 0.5 * \legendTickBottom}
\newcommand\legendXBoxOff{1mm} 
\newcommand\legendXBoxOffBack{4mm}
\newcommand\legendYBoxOff{2.9cm}
\pgfmathsetlengthmacro{\legendXBoxWidth}{0.995*\textwidth}
\newcommand\legendYBoxHeight{4.5mm} 
\newcommand\legendSkip{4mm}
\newcommand\exampleIdx{178}
\newcommand\legendRectEdge{1mm}

\foreach \x / \w / \h / \frameIdx / \i [
    remember=\dw as \lastWidth (initially 0), 
    evaluate=\w as \dw using \w+\lastWidth,
]in {
    0/4.4091/4.0939/0.0/2798,
    1/4.4091/4.0939/1.0/494,
    2/4.4091/4.0939/2.0/849,
    3/4.4091/4.0939/2.5/2409,
    4/4.4091/4.0939/3.0/2798,
    5/4.4091/4.0939/4.0/494,
    6/4.4091/4.0939/5.0/849,
    7/4.4091/4.0939/6.0/849,
    8/4.4091/4.0939/6.5/2409,
    9/4.4091/4.0939/7.0/2409,
    10/4.4091/4.0939/8.0/2409,
    11/4.4091/4.0939/9.0/2409
    }{

    \def\fileName{frame_\frameIdx}
    
    \newcommand\relScale{5.4} 
    \pgfmathsetlengthmacro{\dOff}{\lastWidth / 5.375 * \doaWidth} 
    \pgfmathsetlengthmacro{\W}{\w / \relScale * \doaWidth} 
    \pgfmathsetlengthmacro{\H}{\h / \relScale * \doaWidth} 
    \node[anchor=center] at (\plotStartX + \lrDist + \dOff + \x * \doaDist + 0.5 * \W, \plotStartY + 1*\doaHeight + \tbDist + 0.5 * \H) {%
        \pgfimage[height=\H, width=\W]{images/beampatterns/example_\exampleIdx/\fileName.pdf} 
    };
    \pgfmathsetmacro{\widthNum}{int(\w)}
    \pgfmathsetlengthmacro{\widthOff}{\doaWidth / \relScale}
    \foreach \wn in {0,...,\widthNum}{
        \pgfmathtruncatemacro\tmp{int(\wn/2) * 2}
        \ifnum\tmp=\wn
            \pgfmathsetmacro{\tickLen}{\majorTick}  
            \node[anchor=north] at (\plotStartX  + \lrDist + \dOff + \x * \doaDist + \wn * \widthOff, \plotStartY + \doaHeight + \tbDist) {\tickFont \wn};
        \else
            \pgfmathsetmacro{\tickLen}{\minorTick}
        \fi
        \draw[line width=\tickWidth] (\plotStartX  + \lrDist + \dOff + \x * \doaDist + \wn * \widthOff, \plotStartY + \doaHeight + \tbDist) --  (\plotStartX  + \lrDist + \dOff + \x * \doaDist + \wn * \widthOff, \plotStartY - \tickLen + \doaHeight + \tbDist);
    }
    \pgfmathsetmacro{\heightNum}{int(\h)}
    \pgfmathsetlengthmacro{\heightOff}{\doaWidth / \relScale}
    \foreach \hn in {0,...,\heightNum}{
        \pgfmathtruncatemacro\tmp{int(\hn/2) * 2}
        \ifnum\tmp=\hn
            \pgfmathsetmacro{\tickLen}{\majorTick}
            \ifthenelse{\x = 0}{
                \node[anchor=east] at (\plotStartX  + \lrDist + \dOff + \x * \doaDist, \plotStartY + \doaHeight + \tbDist + \hn * \heightOff) {\tickFont \hn};
            }{}
        \else
            \pgfmathsetmacro{\tickLen}{\minorTick}
        \fi
        \draw[line width=\tickWidth] (\plotStartX  + \lrDist + \dOff + \x * \doaDist - \tickLen, \plotStartY + \doaHeight + \tbDist + \hn * \heightOff) --  (\plotStartX  + \lrDist + \dOff + \x * \doaDist, \plotStartY + \doaHeight + \tbDist + \hn * \heightOff);
    }

    \pgfmathtruncatemacro{\frameInt}{\frameIdx}
    \pgfmathsetmacro{\timeval}{\frameInt /20}
    \node[anchor=center, xshift=\tXoff, yshift=\tYoff] at (\plotStartX + \lrDist + \dOff + \x * \doaDist + 0.5 * \W, \plotStartY + 1*\doaHeight + \tbDist + 0.5 * \H) {\scriptsize \contour{white}{\frameIdx\,s}};
    
}

\node[anchor=north] at (0.5 * \linewidth + 2mm, \plotStartY- 3*\majorTick + \doaHeight + \tbDist-0.25mm) {\footnotesize room width [m]};
\node[anchor=south, rotate=90] at (\plotStartX  + \lrDist - 2mm, \plotStartY + \doaHeight+ 0.5 * \doaHeight + \tbDist + 1mm) {\footnotesize length [m]};

\newcommand\labelColor{tab_orange}
\newcommand\legendDeltaX{4mm}
\draw[draw, rounded corners=\legendRectEdge, line width=\tickWidth]
            (\plotStartX + \legendXBoxOffBack, \plotStartY + \legendYBoxOff) rectangle (\plotStartX + \legendXBoxWidth + \legendXBoxOffBack, \plotStartY + \legendYBoxOff + \legendYBoxHeight);

\pgfmathsetlengthmacro{\micLegendDist}{
 4mm+\legendDeltaX
 } 
\node[anchor=west] at (\plotStartX + \legendXBoxOff + \micLegendDist, \plotStartY + \legendYBoxOff+ 0.5 * \legendYBoxHeight) {\footnotesize mic.\,array};
\node[star,star points=5,draw,fill=black,minimum size=5pt,star point ratio=2.25, anchor=center, inner sep=0pt, rounded corners=0.1pt, xshift=-1mm] at(\plotStartX + \legendXBoxOff + \micLegendDist, \plotStartY + \legendYBoxOff+ 0.5 * \legendYBoxHeight) {};
            
 \pgfmathsetlengthmacro{\oracleDist}{
 2cm+2*\legendDeltaX
 } 
 \newcommand\yRoomOff{0mm}
 \draw[color=\labelColor, line width=1.5pt,opacity=0.5,decorate,
  decoration={snake, amplitude=3pt, segment length=10pt}] (\plotStartX + \legendXBoxOff + \oracleDist, \plotStartY + \legendYBoxOff+ 0.5 * \legendYBoxHeight + \yRoomOff) -- coordinate (snakeend) (\plotStartX + \legendXBoxOff + \oracleDist+ \legendSkip, \plotStartY + \legendYBoxOff+ 0.5 * \legendYBoxHeight + \yRoomOff);
   \draw (\plotStartX + \legendXBoxOff + \oracleDist, \plotStartY + \legendYBoxOff+ 0.5 * \legendYBoxHeight) 
node[cross=3.5pt, color=\labelColor, line width=1.75pt, anchor=center, rotate=45, opacity=0.5]{};
  \fill[\labelColor] (\plotStartX + \legendXBoxOff + \oracleDist+ \legendSkip, \plotStartY + \legendYBoxOff+ 0.5 * \legendYBoxHeight + \yRoomOff) circle (2pt);
 \node[anchor=west, xshift=0.5mm] at (\plotStartX + \legendXBoxOff + \oracleDist+ \legendSkip, \plotStartY + \legendYBoxOff+ 0.5 * \legendYBoxHeight){\footnotesize trajectory from start ({\protect\tikz[baseline=-0.6ex] 
\node[
    cross=3.5pt,
    color=\labelColor,
    line width=1.75pt,
    rotate=45,
    opacity=1.0,
    anchor=center
] {};
}\hspace{-0.6mm}) to current speaker pos. ({\protect\tikz \protect\fill[\labelColor] (0mm,0mm) ++ (0mm,0.75mm) circle[radius=1.75pt];})};

 \pgfmathsetlengthmacro{\drivingDist}{
 9cm+3*\legendDeltaX
 } 
 \node at (\plotStartX + \legendXBoxOff + \drivingDist, \plotStartY + \legendYBoxOff+ 0.5 * \legendYBoxHeight) 
 {\pgfimage[height=3.5mm]{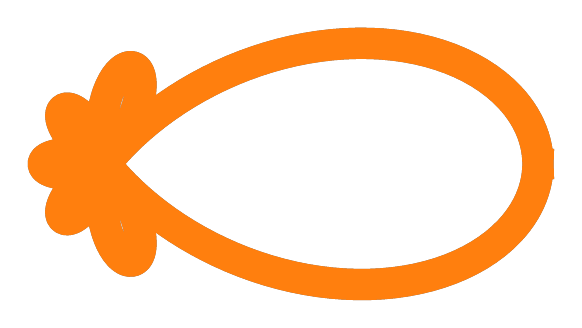} };
 \node[xshift=3mm, anchor=west] at (\plotStartX + \legendXBoxOff + \drivingDist, \plotStartY + \legendYBoxOff+ 0.5 * \legendYBoxHeight) {
 	\footnotesize broadband beampattern $\sum_k |\mathbf{w}_{tk}^\hermitian \mathbf{d}(\theta, \phi) |^2$, peak normalized
};

\end{tikzpicture}
\vspace*{-20pt}
\caption{
Time-varying spatial selectivity of mask-based \ac{mvdr} beamforming for moving speaker extraction with third-order ambisonics.
The simulated two speaker ({\protect\tikz[baseline=-1.0ex] 
  \protect\fill[tab_blue] (0,0) circle (1.5pt);}/\hspace{-0.8mm} {\protect\tikz[baseline=-0.4ex] 
  \protect\fill[tab_orange] (0,0) circle (1.5pt);}) trajectories follow the modified social force motion model from~\cite{kienegger26ar_guided_ssf_journal}. Further visualization are available online\textsuperscript{\ref{project_page}}.
}
\label{fig:beampattern}
\vspace*{-10pt}
\end{figure*}

\subsection{Weakly guided neural mask estimation}

For neural mask estimation, we propose to decouple temporal-spectral from spatial processing, by using the sound field power $\lVert \mathbf{X}_{tk} \rVert^2$ instead of stacked ambisonics coefficients as \ac{dnn} input.
On top of an ambisonics-order-agnostic architecture, without spatial information, the resulting mask estimator is unaffected by closely spaced speakers, since it cannot exploit source location.
The speech and noise masks can be learned implicitly via a signal reconstruction loss~\cite{boeddeker21ctf_e2e_training,ochiai23moving_speaker_attention_mvdr}, however, this reduces interpretability and becomes ambiguous during speaker crossings under a distortionless constraint.
We thus use an \ac{irm} as training target, defined as
\begin{equation} \label{eq:irm}
\sqrt{\mathcal{M}^\superscript{(\mathrm{\speechNoise})}_{tk}} = 
    \begin{cases} 
        {\lVert \speechNoiseBold_{tk}\rVert}^2 \Big/ \big({{\lVert \speechNoiseBold_{tk}\rVert}^2 + {\lVert \mathbf{X}_{tk}\!-\!\speechNoiseBold_{tk}\rVert}^2}\big), &  \lVert \speechNoiseBold_{tk}\rVert > \mathcal{E}_{\mathcal{M}}^\superscript{(\mathrm{\speechNoise})} ,\\ 
        0, & \text{else,}
    \end{cases}  
\end{equation} 
with $\mathbf{\speechNoiseBold}_{tk} \in \{\mathbf{S}_{tk}, \mathbf{V}_{tk}\}$ and threshold $\mathcal{E}_{\mathcal{M}}^\superscript{(\mathrm{\speechNoise})}$ to avoid covariance contamination during periods of silence.
Similar to \cite{zhang24noise_covariance_estimation_irm_definition}, one can show that for $\lVert \speechNoiseBold_{tk}\rVert > \mathcal{E}_{\mathcal{M}}^\superscript{(\mathrm{\speechNoise})}$, $\mathcal{M}^\superscript{(\mathrm{\speechNoise})}_{tk}$ minimizes the Frobenius norm of the \mbox{rank-1} covariance update error ${\lVert \mathcal{M}^\superscript{(\mathrm{\speechNoise})}_{tk} \mathbf{X}_{tk}\mathbf{X}_{tk}^\hermitian - \speechNoiseBold_{tk}\speechNoiseBold_{tk}^\hermitian \rVert}_{F}$ under the condition $\mathbf{S}_{tk}\perp\mathbf{V}_{tk}$.
While the latter is not valid in general, especially not for closely spaced and crossing speakers, it provides a bounded training target, which solely depends on signal power.
Furthermore, since $\sqrt{\mathcal{M}^\superscript{(\mathrm{S})}_{tk}}  +\sqrt{\mathcal{M}^\superscript{(\mathrm{V})}_{tk}} = 1$ only holds in the restrictive case of $\mathcal{E}_{\mathcal{M}}^\superscript{(\mathrm{S})} = \mathcal{E}_{\mathcal{M}}^\superscript{(\mathrm{V})} = 0$, both speech and noise masks are estimated separately in this work.
However, the input power $\lVert \mathbf{X}_{tk} \rVert^2$ alone provides insufficient information 
to distinguish
target and interfering speakers, which is necessary for mask estimation. 
We therefore complement $\lVert \mathbf{X}_{tk} \rVert^2$ by additional input features based on the target's starting direction, resulting in \textit{weakly guided} mask estimation.

\textbf{\textit{Fixed beamformer}} Using a beamformer oriented towards the target speaker as \ac{dnn} conditioning has been extensively studied for stationary~\cite{nugraha16source_separation_das_input,perotin18lstm_ambisonics_beamforming,elminshawi23bf_guided_tse} and also recently for moving speakers~\cite{iatariene25tracking_speaker_embeddings}. 
We propose to adapt this concept to the weakly guided scenario, by using a \ac{fb} oriented towards the target's starting direction.
This gives the neural mask estimator the opportunity to learn the target's temporal-spectral characteristic at recording start, until the target moves out of the beam width.
To parameterize the \ac{fb}, we re-use the initialization scheme in \cref{eq:init_covaraince}, as shown in  \cref{fig:overview}. 

\textbf{\textit{Autoregression}}
A frame-wise causal, sequential processing style allows leveraging previously enhanced frames via temporal feedback during the estimation of the current frame.
In our previous work, we could demonstrate how integrating this additional speaker-specific cue improves enhancement performance of deep, non-linear spatial filters~\cite{kienegger25deep_joint_ar_ssf}.
Here we propose utilizing autoregression for mask estimation by including the processed speech $\hat{S}_{\delayIdx{t}k}$ as additional \ac{dnn} input at current frame $t$.
Compared with the \ac{fb}, the \ac{ar} feedback provides continuous target speaker information, at the price of a constant single frame delay.
Optionally, both \ac{fb} and \ac{ar} input features can be also used jointly, as illustrated in \cref{fig:overview}.

\section{Experimental Setup}
\label{sec:experiments}
\subsection{Dataset}\label{sec:dataset}
\hspace{\the\parindent}\textit{\textbf{Synthetic dataset}}
To enable development and evaluation in a controlled acoustic scenario,
we create a synthetic dataset containing noisy and reverberant two-speaker mixtures. 
In particular, we follow the simulation setup from~\cite{kienegger26ar_guided_ssf_journal} and spatialize paired utterances from Libri2Mix corresponding to speaker trajectories obtained by the social force motion model~\cite{helbing95social_force_model}, see \cref{fig:beampattern}.
We generate third-order ambisonics coefficients using a custom implementation of the image method based on an idealized plane-wave model under far-field assumptions ~[\citen{rafaely19spherical_array_processing}, Sec.~2.5].
To account for the frequency-dependency induced by regularized mode strength compensation, i.e. inverting $b_n(\kappa_k r_p)$ in \cref{eq:ambisonics_expansion}, we assume a rigid spherical array matching the dimensions of an MH Acoustics \texttt{Eigenmike64}, and apply its Tikhonov-regularized response (15\,dB max gain) in the \ac{stft} domain~\cite{mccormack18array2sh_reference}.
For the latter, we use a $\sqrt{\text{Hann}}$ window of length $32\,\mathrm{ms}$ and $16\,\mathrm{ms}$ hop-size.
Finally, we add sensor noise at -30\,dB \ac{snr}, which is amplified at low frequencies in the ambisonics domain due mode strength compensation ~\cite{moreau06hoa_4th_order}.

\textit{\textbf{Recorded dataset}}
To assess generalizability and robustness in real-world scenarios, we also include recordings with human speakers for evaluation.
We use an MH Acoustics \texttt{Eigenmike64} as recording device, which we place centered in a meeting room with a reverberation time of approximately 500\,ms.
In each recording, two male, non-native English speakers simultaneously read out segments from the Rainbow Passage \cite{fairbanks60rainbow_passage}. 
While one speaker remains seated and stationary, the other stands up, walks past the seated speaker to the opposite side of the table, and sits down again, thereby including a directional speaker crossing in each recording.
We split the Rainbow Passage into 3 segments of roughly equal length and permute these among both speakers, totaling to 6 recordings of about 30\,s each.
Videos of the recordings can be found on our project page\footnote{\label{project_page}\projectPage}.

\subsection{Model and algorithm parameterization}\label{sec:model_parameterization}
\hspace{\parindent}\textit{\textbf{Mask estimation}}
Weakly guided mask estimation requires the exploitation of temporal-spectral correlations between recorded sound field and speaker-specific input features.
We therefore choose the convolutional-recurrent architecture CRUSE~\cite{braun21cruse} as a low-complexity baseline, which has proven efficient for both single and multichannel speech enhancement~\cite{kindt22sep}.
We adapt CRUSE for frame-wise causal processing by using padded CNN and unidirectional GRU layers, yielding 2.3\,GMACs$/$s at 6.9\,M parameters.
Additionally, we employ \mbox{SpatialNet}~\cite{quan24spatialnet}, a state-of-the-art  \ac{dnn} architecture using state-space modeling for efficient and streamable multichannel speech enhancement.
In our configuration, \mbox{SpatialNet} requires 18.7\,GMACs$/$s at 1.7\,M parameters.
For both architectures, using \ac{fb} and \ac{ar} input features simultaneously results in a negligible increase of 500 parameters and less than 10\,MMACs$/$s.

\textit{\textbf{MVDR beamformer}} 
The \ac{mvdr} beamformer requires an inversion of the noise covariance matrix, for which we use the Cholesky decomposition. 
For improved conditioning, we employ Tikhonov regularization---in the beamforming literature commonly referred to as diagonal loading---which corresponds to a diffuse noise prior in the ambisonics domain.
In dynamic scenarios, directionally crossing speakers cause a singularity in the \ac{mvdr} beamformer formulation, since the inverted noise covariance matrix approaches a projector orthogonal to the target steering vector.
As a result, both numerator and denominator in \cref{eq:miso_mvdr} jointly vanish, which evaluates to a unit gain mathematically.
In practice, we artificially enforce this behavior using a thresholded pass-through for numerical stability. 
Regarding the \ac{pev}-based steering vector computation, we employ five Power Iteration steps with additional pre-whitening to reduce noise contamination from the soft-masking based covariance matrix estimate~\cite{boeddeker21ctf_e2e_training}.
We note that computational efficiency could be further increased by using recursive eigenvector tracking, see e.g.~\cite{zaidel26rtf_binaural_beamforming_moving_speaker}. However, this lies outside the scope of this work.
 
\textit{\textbf{Training and optimization strategy}}
All non-learnable hyperparameters used in beamforming,  covariance estimation and \ac{irm} definition are optimized via an exhaustive search on the validation dataset.
To train the neural mask estimators, we employ an $\ell_1$ loss with an initial learning rate of 10$^{-3}$ and exponential decay, which decimates the learning rate over the total training time of 50 epochs.
To avoid the inherent non-parallelizability of the \ac{ar} methods, we use \acl{rds} as pseudo-\ac{ar} training strategy \cite{shen25arise}.

\section{Results}
\label{sec:results}
\newcommand\tableW{3pt}
\newcommand\tableWL{8pt}
\newcommand\tableHspacer{-0pt} %
\newcommand\tableHrow{-0pt} %

\def\tableW{3pt} %
\begin{table}[t!]
 \caption{
 Mask estimation methods for weakly guided speaker extraction based on fixed beamforming (FB) and autoregression (AR).
}
\vspace{-5pt}
\label{tab:results}
\resizebox{\linewidth}{!}{
\renewcommand{\arraystretch}{0.9}
\footnotesize
\begin{tabular}{
c@{\hspace{6pt}}l@{\hspace{5pt}}c@{\hspace{\tableW}}c@{\hspace{\tableW}}c@{\hspace{5pt}}c@{\hspace{\tableW}}c@{\hspace{\tableW}}c@{\hspace{\tableW}}c %
}
 \toprule[1.5pt]
 & \multicolumn{4}{c}{\textbf{Mask Estimation Method}} & \multicolumn{3}{c}{\textbf{Enhancement Performance}} \\[1pt]
 \cmidrule(l{0pt}r){2-5} \cmidrule(l{0pt}r){6-8} 
\textbf{ID} & Architecture & FB & AR & MACs\,[G/s] &  \acs{pesq}\,$\uparrow$ & \acs{estoi}\,[\%]\,$\uparrow$ & \acs{wer}\,[\%]\,$\downarrow$ \\ \midrule
\picTable{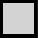}  & Unprocessed & $-$ & $-$ & $-$ & 1.13\myPM.01 & 43.4\myPM.3 & \hspace*{-1.0mm}112.3\myPM2.4 \\[-1pt] \cmidrule(lr){1-8}
\picTable{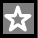} & IRM\,(oracle) & $-$ & $-$ & $-$ & 1.93\myPM .01 & 77.8\myPM.3 & 10.7\myPM0.5 \\[-1pt] \cmidrule(lr){1-8}
\picTable{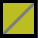}  & CRUSE$/$SpatialNet & \textcolor{green}{\ding{51}} & \textcolor{red}{\ding{55}} & 2.3$/$18.7 & 1.38$/$1.71 & 59.2$/$71.1 & 45.4$/$17.5 \\
\picTable{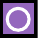}  & CRUSE$/$SpatialNet & \textcolor{red}{\ding{55}} & \textcolor{green}{\ding{51}} & 2.3$/$18.7 &1.60$/$1.70 & 68.3$/$71.2 & 24.0$/$18.0 \\
\picTable{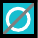}  & CRUSE$/$SpatialNet & \textcolor{green}{\ding{51}}  & \textcolor{green}{\ding{51}}  & 2.3$/$18.7 & \textbf{1.65}$/$\textbf{1.77} & \textbf{69.2}$/$\textbf{73.3} & \textbf{22.5}$/$\textbf{15.5} \\ 
 \bottomrule[1.5pt]
\end{tabular}
}

\vspace{1mm}
\footnotesize

\begin{tabular}{>{\setlength{\leftskip}{0.2em}}p{0.95\columnwidth}}
We report the sample means with 95\% confidence intervals, which are $\myPM.01$\,PESQ, $\myPM.3$\%\,ESTOI and $\myPM1$\%\,WER for the \acp{dnn} (\picLegend{images/mvdr_beamformer/hatch_legend/fb_ambix.pdf},\,\picLegend{images/mvdr_beamformer/hatch_legend/ar_ambix.pdf},\,\picLegend{images/mvdr_beamformer/hatch_legend/dual_ambix.pdf}).
\end{tabular}
\vspace{-10pt}
\end{table}

\begin{figure}[b!]
\centering
\subfloat[Influence of target duration close to initial azimuth angle $\theta_0$.\label{fig:azimuth_distance}]{
  \begin{minipage}[b]{0.48\linewidth} %
    \vspace{0pt}
    \centering
\hspace*{-2mm}\begin{tikzpicture}

\newcommand\figSize{1.7cm} 
\pgfmathsetlengthmacro{\figWidth}{\figSize * 3.0 / 3.0}
\pgfmathsetlengthmacro{\figDist}{1.5mm} 
\pgfmathsetlengthmacro{\figHeight}{\figSize * 1.0}
\newcommand\plotStartX{0mm}
\newcommand\plotStartY{0mm}
\colorlet{trainingColor}{red!70}
\colorlet{testColor}{blue!70}
\newcommand\rowSplitSkip{-0.75mm}

\newcommand{\misoRef}{PEV \cref{eq:miso_mvdr}}
\newcommand{\mimoRef}{Ref \cref{eq:mimo_mvdr}}

\foreach \plotIdx/\azimuthThreshold in {%
    0/10,%
    1/20%
}{
    \node[anchor=center] at (\plotStartX + 0.5 * \figWidth + \plotIdx * \figWidth + \plotIdx * \figDist, \plotStartY + 0.5 * \figHeight) {%
        \pgfimage[height=\figHeight, width=\figWidth]{images/azimuth_distance/recording_length_pesq_\azimuthThreshold.pdf} 
    };


     \ifnum\plotIdx=0
        \def\numTicks{8}
    \else
        \def\numTicks{8}
    \fi
    \pgfmathsetlengthmacro{\labelDist}{
            \figHeight / \numTicks
        }
    \foreach \y in {%
        0,...,\numTicks%
    } {%
        \pgfmathtruncatemacro{\tmp}{int(\y / 2) * 2}
            \ifnum\tmp=\y
            \ifnum\y=0
                 \pgfmathsetmacro{\yLabel}{1.6}
            \else
                \ifnum\y=2
                     \pgfmathsetmacro{\yLabel}{1.65}
                \else
                    \ifnum\y=4
                         \pgfmathsetmacro{\yLabel}{1.7}
                         \else
                         \ifnum\y=6
                         \pgfmathsetmacro{\yLabel}{1.75}
                        \else
                            \pgfmathsetmacro{\yLabel}{1.8}
            \fi
            \fi
            \fi
            \fi
                \def\tickLen{\majorTick}
                \ifnum\plotIdx=0
                    \node[anchor=east] at (\plotStartX + \plotIdx * \figWidth + \plotIdx * \figDist,\plotStartY + \y * \labelDist) {\tickFont \yLabel};
                \fi
            \else
                \def\tickLen{\minorTick}
            \fi
                
                
        
        \draw[line width=\tickWidth] (\plotStartX + \plotIdx * \figWidth + \plotIdx * \figDist,\plotStartY + \y * \labelDist) -- (\plotStartX +\plotIdx * \figWidth + \plotIdx * \figDist- \tickLen,\plotStartY + \y * \labelDist);
    }
    \ifnum\plotIdx=0
    \node[anchor=south, rotate=90, yshift=4.75mm] at (\plotStartX + \plotIdx * \figWidth + \plotIdx * \figDist,\plotStartY + 0.5 * \figHeight) {\footnotesize PESQ\,$\rightarrow$};
    \fi

    \foreach \xNum / \xRel / \xLabel in {
        0/0.2/{[0,\,1)},
        1/0.5/??,
        2/0.8/{[2,\,3)}
    } {
    \ifnum\xNum=1
    
    \else
        \node[anchor=north, rotate=0, yshift=0mm, xshift=0.5mm] at (\plotStartX + \xRel * \figWidth + \plotIdx * \figWidth + \plotIdx * \figDist,\plotStartY) {\tickFont \xLabel};
    \fi
        \draw[line width=\tickWidth] (\plotStartX + \xRel * \figWidth + \plotIdx * \figWidth + \plotIdx * \figDist,\plotStartY) -- (\plotStartX + \xRel * \figWidth+ \plotIdx * \figWidth + \plotIdx * \figDist,\plotStartY- \majorTick);
    }

    \node[xshift=6mm, yshift=0.5mm] at (\plotStartX + \plotIdx * \figWidth + \plotIdx * \figDist,\plotStartY + 0.9 * \figHeight) {\footnotesize $\Delta \theta\!=\!{\azimuthThreshold}^\circ$};
}

\node[yshift=-6mm, xshift=-4mm] at (\plotStartX + 1 * \figWidth + 0.5 * \figDist,\plotStartY) {time [s] until $|\theta_t - \theta_0| > \Delta\theta$};

\end{tikzpicture} %
\vspace{-15pt}
  \end{minipage}
}
\hfill
\subfloat[Long-form recordings using \texttt{freeform} speech from EARS.]{
  \begin{minipage}[b]{0.46\linewidth} %
    \vspace{0pt}
    \centering
    \hspace*{-1mm}\begin{tikzpicture}
    
\newcommand\doaWidth{1.5cm}
\newcommand\doaHeight{8mm}
\newcommand\figSize{2.075cm} 
\newcommand\labelFont{\footnotesize}
\pgfmathsetlengthmacro{\figWidth}{\figSize * 1.5}
\pgfmathsetlengthmacro{\figHeight}{\figSize * 0.9}
\newcommand\labelOff{5.5mm} 
\newcommand\doaDist{1cm} 
\newcommand\doaTickVoff{-0.2mm}
\newcommand\doaTickHoff{-0.15mm}
\newcommand\doaLabelDist{3mm}
\newcommand\lrDist{10mm}
\newcommand\tbDist{8mm}
\newcommand\plotStartX{0mm}
\newcommand\plotStartY{0mm}
\newcommand\tickSize{7pt}
\newcommand\tickSkip{9pt}
\newcommand\timeDist{3.1mm} %
\newcommand\legendTick{4mm}
\newcommand\legendTickTop{-0.25mm}
\newcommand\legendTickBottom{-0.5mm}
\pgfmathsetlengthmacro{\legendTickMiddle}{0.5 * \legendTickTop + 0.5 * \legendTickBottom}
\newcommand\legendOff{1mm}
\newcommand\legendYOff{-2mm} %
\newcommand\centerOff{-1.1cm}
\newcommand\legendWidth{4cm}
\newcommand\legendSkip{4mm}
\newcommand\legendOffProp{2.7cm}
\newcommand\labelWidth{1pt}
\newcommand\descriptX{2.7mm}
\newcommand\descriptY{2.2mm}
\newcommand\legendXBoxOff{1.25cm}
\newcommand\legendYBoxOff{4mm}
\newcommand\legendXBoxWidth{2.8cm}
\newcommand\legendYBoxHeight{2.05cm}

\foreach \x / \low / \high / \metric / \numTicks / \metricLabel / \relWidth / \method
[
    remember=\dr as \lastRel (initially 0), 
    evaluate=\relWidth as \dr using \relWidth+\lastRel,
]
in {
    0/4/10/snr/6/{\small $\Delta$\,SNR\,[dB]\,$\rightarrow$}/1.5/spatialnet_snr
    }{
    
    \pgfmathsetlengthmacro{\figWidth}{\figSize * \relWidth}
    \node[anchor=center] at (\plotStartX + \x * \doaDist + \lastRel * \figSize, \plotStartY + 0.5 * \figHeight) {%
        \pgfimage[height=\figHeight, width=\figWidth]{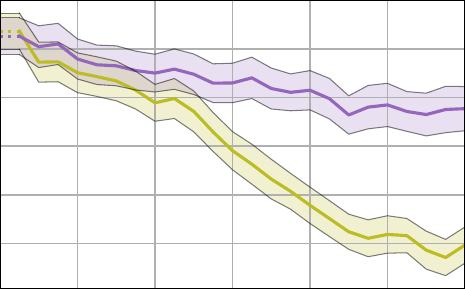} 
    };
    \def\xTickNum{6}
    \pgfmathsetlengthmacro{\xTickW}{
    \figWidth / (\xTickNum)
    }
    \foreach \xTick in {0, ..., \xTickNum}{
        \draw[line width=\tickWidth] (\plotStartX + \x * \doaDist + \lastRel * \figSize + \xTick * \xTickW - \figWidth/2, \plotStartY) -- (\plotStartX + \x * \doaDist + \lastRel * \figSize + \xTick * \xTickW - \figWidth/2, \plotStartY - \majorTick);
        \pgfmathsetmacro{\xTickLabel}{int(10 * \xTick)}
        \node[anchor=north] at (\plotStartX + \x * \doaDist + \lastRel * \figSize + \xTick * \xTickW - \figWidth/2, \plotStartY) {\labelFont\xTickLabel};
    }
    \node[yshift=-5mm] at (\plotStartX + \x * \doaDist + \lastRel * \figSize, \plotStartY) {time [s]};

    \pgfmathsetlengthmacro{\deltaY}{\figHeight / \numTicks}
    \foreach \y in {0, ..., \numTicks}{
        \ifthenelse{\y = \numTicks}{
                \newcommand\metricTick{\high}
            }{
                \pgfmathsetmacro{\metricTick}{
            int(
                (\low + (\high - \low) / \numTicks * \y
            )
            }
        }
        \pgfmathtruncatemacro\tmp{int(\y/2) * 2}
        \ifnum\tmp=\y
            \pgfmathsetmacro{\tickLen}{\majorTick}
            \node[anchor=east] at (\plotStartX + \lastRel * \figSize + \x * \doaDist  - 0.5 * \figWidth,\plotStartY + \y * \deltaY) {\labelFont \metricTick};
        \else 
            \pgfmathsetmacro{\tickLen}{\minorTick}
        \fi
        
        \ifthenelse{\equal{\metric}{mae}}{
        \ifnum\y=0
\node[anchor=center, rotate=90] at (\plotStartX + \lastRel * \figSize + \x * \doaDist - 0.5 * \figWidth - \labelOff + 2mm,\plotStartY + 0.5 * \figHeight) {\metricLabel};
        \fi        
        \pgfmathsetmacro{\metricTick}{int((\high - \low) / \numTicks * \y)}
         
            \node[anchor=east] at (\plotStartX + \lastRel * \figSize + \x * \doaDist - \tickLen - 0.5 * \figWidth,\plotStartY + \y * \deltaY) {\labelFont \metricTick};
        }{
            \ifnum\y=0
            \node[anchor=center, rotate=90] at (\plotStartX + \lastRel * \figSize + \x * \doaDist - 0.5 * \figWidth - \labelOff + 0.5mm,\plotStartY + 0.5 * \figHeight) {\small \metricLabel};
            \fi
        }
        \draw[line width=\tickWidth] (\plotStartX + \lastRel * \figSize + \x * \doaDist - \tickLen - 0.5 * \figWidth,\plotStartY + \y * \deltaY) --  (\plotStartX + \lastRel * \figSize + \x * \doaDist - 0.5 * \figWidth,\plotStartY + \y * \deltaY);
    }
}

\end{tikzpicture}
    \label{fig:long_form_audio}
    \vspace{-16pt} %
  \end{minipage}
}

\vspace{-5pt}
\caption{Fixed beamformer (FB,\,\picLegend{images/mvdr_beamformer/hatch_legend/fb_ambix.pdf}) vs. autoregressive (AR,\,\picLegend{images/mvdr_beamformer/hatch_legend/ar_ambix.pdf}) conditioning for weakly guided \ac{mvdr} parameterization with \mbox{SpatialNet}.}
\label{fig:performance_synthetic}
\vspace{0pt}
\end{figure}

We use our synthetic dataset for a detailed \textit{analysis} and lab recordings to assess real-world \textit{generalizability}. 
With the synthetic dataset availing ground truth speech signals, we employ the intrusive metrics \acs{pesq} and \acs{estoi} as measures for speech quality and intelligibility, respectively. 
Additionally, we leverage the transcription of a downstream \ac{asr} system to compute the \ac{wer} relative to known reference segments, which enables the evaluation of our real-world meeting recordings.
In particular, we utilize the lightweight \ac{asr} model \texttt{QuartzNet15x5Base-En}~\cite{kuchaiev9nemo}, which is trained solely on clean and telephony conversational English speech. 
Since this increases sensitivity to signal distortions, the \ac{wer} provides a joint measure for processing degradation as well as task-level performance. 

\textbf{\textit{Synthetic dataset}}
\Cref{tab:results} presents the enhancement results with CRUSE and SpatialNet as neural mask estimators for \ac{mvdr} beamformer parameterization. 
For CRUSE, using the \ac{fb} as input feature consistently underperforms across all metrics~(\picLegend{images/mvdr_beamformer/hatch_legend/fb_ambix.pdf}).
Due to its limited computational capacity, CRUSE is unable to  exploit the target speaker characteristics provided by the \ac{fb} sufficiently. 
On the other hand, the \ac{ar} feedback from the beamformer output considerably improves enhancement (\picLegend{images/mvdr_beamformer/hatch_legend/ar_ambix.pdf}), proving its capability to compensate for the limited guidance from the initial \ac{doa}. 
Due to the increased complexity and superior sequential modeling capability via a state-space based architecture, SpatialNet performs similarly for both \ac{fb} and \ac{ar} input features across the test set.
While the former provides a temporally aligned cue, it relies on implicitly extracting an informative speaker embedding at recording start.
This depends on the representational capacity of the \ac{dnn}, but also on the duration of the target within proximity of the starting direction, as shown in \cref{fig:azimuth_distance}. 
Therefore, the \ac{fb} feature is highly sensitive to speaker movement at recording start~(\picLegend{images/mvdr_beamformer/hatch_legend/fb_ambix.pdf}), while \ac{ar} guidance retains consistent enhancement across varying motion dynamics~(\picLegend{images/mvdr_beamformer/hatch_legend/ar_ambix.pdf}).
To evaluate the long-form audio processing capability, we generate an additional test set using all \texttt{freeform} speech utterances from the EARS corpus~\cite{richter24ears}.
In particular, we compute the \ac{snr} on the first minute of the 322 synthetic two-speaker mixtures in 2.5\,s intervals. 
\Cref{fig:long_form_audio} demonstrates how after the first half of the recording, SpatialNet guided by the \ac{fb} (\picLegend{images/mvdr_beamformer/hatch_legend/fb_ambix.pdf}) drops in performance while autoregression  maintains consistent noise reduction~(\picLegend{images/mvdr_beamformer/hatch_legend/ar_ambix.pdf}). 
Conclusively, the \ac{fb} and \ac{ar} input features represent a \textit{tradeoff} between temporal cue alignment, which is beneficial for short recordings, and generalizability across speaker movement and recording duration.
As a result, their \textit{combination} yields superior enhancement for both CRUSE and SpatialNet~(\picLegend{images/mvdr_beamformer/hatch_legend/dual_ambix.pdf}). 

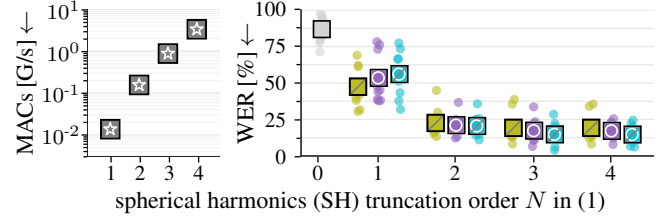
\begin{figure}[t!]
\begin{tikzpicture}

\newcommand\doaWidth{1.5cm} 
\newcommand\doaHeight{8mm}
\newcommand\figSize{1.95cm} 
\pgfmathsetlengthmacro{\figWidth}{\figSize * 2.2}
\pgfmathsetlengthmacro{\figWidthOne}{\figSize * 0.9}
\pgfmathsetlengthmacro{\figWidthTwo}{\figSize * 2.4} 
\pgfmathsetlengthmacro{\figHeight}{\figSize}
\newcommand\labelOff{7mm} 
\newcommand\labelFont{\footnotesize}
\newcommand\doaDist{-4mm} 
\newcommand\doaTickVoff{-0.2mm}
\newcommand\doaTickHoff{-0.15mm}
\newcommand\doaLabelDist{3mm}
\newcommand\lrDist{10mm}
\newcommand\tbDist{8mm}
\newcommand\plotStartX{0mm}
\newcommand\plotStartY{0mm}
\newcommand\tickSize{7pt}
\newcommand\tickSkip{9pt}
\newcommand\timeDist{3.1mm} %
\newcommand\legendTick{4mm}
\newcommand\legendTickTop{-0.25mm}
\newcommand\legendTickBottom{-0.5mm}
\pgfmathsetlengthmacro{\legendTickMiddle}{0.5 * \legendTickTop + 0.5 * \legendTickBottom}
\newcommand\legendOff{0.5mm} 
\newcommand\legendYOff{-2mm} %
\newcommand\centerOff{-1.1cm}
\newcommand\legendWidth{4cm}
\newcommand\legendSkip{4mm}
\newcommand\legendOffProp{2.7cm}
\newcommand\labelWidth{1pt}
\newcommand\descriptX{2.7mm}
\newcommand\descriptY{2.2mm}
\newcommand\legendXBoxOff{1.25cm}
\newcommand\legendYBoxOff{4mm}
\newcommand\legendXBoxWidth{2.8cm}
\newcommand\legendYBoxHeight{2.05cm}

\foreach \x / \low / \high / \l / \metric / \numTicks / \metricLabel [
    remember=\dl as \lastLen (initially 0), 
    evaluate=\l as \dl using \l+\lastLen,
]in 
{
    1/1.0/4.5/7.85/macs/7/{MACs\,[{G/s}]\,$\leftarrow$},
    2/0/1.0/6.45/wer/8/{WER\,[{\%}]\,$\leftarrow$}
    }
    {
    \edef\fileName{\metric}

    \ifnum\x=1
    \pgfmathsetlengthmacro{\figOff}{0.5 * \figWidthOne}
        \node[anchor=center] at (\plotStartX + \figWidthOne + \x * \doaDist, \plotStartY + 0.5 * \figHeight) {%
            \pgfimage[height=\figHeight, width=\figWidthOne]{images/recordings/\fileName.pdf} 
        };
    \else
    \pgfmathsetlengthmacro{\figOff}{\figWidthOne + 0.5 * \figWidthTwo}
        \node[anchor=center] at (\plotStartX +  \figWidthOne + \figWidthTwo + \x * \doaDist, \plotStartY + 0.5 * \figHeight) {%
            \pgfimage[height=\figHeight, width=\figWidthTwo]{images/recordings/\fileName.pdf} 
        };
    \fi

    \ifnum\x=1
        \node[anchor=center, rotate=90, yshift=1mm] at (\plotStartX + \figOff + \x * \doaDist - \labelOff,\plotStartY + 0.5 * \figHeight) {\small \metricLabel};
    \else
        \node[anchor=center, rotate=90, yshift=0mm] at (\plotStartX + \figOff + \x * \doaDist - \labelOff,\plotStartY + 0.5 * \figHeight) {\small \metricLabel};
    \fi
    \ifnum\x=1
        \foreach \exp / \tick in {
            0/3,0/4,0/5,0/6,0/7,0/8,0/9,
1/1,1/2,1/3,1/4,1/5,1/6,1/7,1/8,1/9,
2/1,2/2,2/3,2/4,2/5,2/6,2/7,2/8,2/9,
3/1,3/2,3/3,3/4,3/5,3/6,3/7,3/8,3/9,
4/1
        }{
            \pgfmathsetmacro{\tickYMulti}{
                (log10(\tick) + \exp-1 + 1 - log10(3)) / (2.6 + log10(2.6) + 1 - log10(3))
            }
            \pgfmathsetlengthmacro{\deltaY}{
              \tickYMulti * \figHeight
            }
            \ifthenelse{\tick = 1}{
                \def\tickLen{\majorTick}
                \pgfmathsetmacro{\shiftExp}{int(abs(\exp-3))}
                \ifnum\exp=1
                    \node[anchor=east, xshift=1mm] at (\plotStartX + \figOff + \x * \doaDist - \tickLen,\plotStartY + \deltaY) {\footnotesize $10^{\text{-}\shiftExp}$};
                \else
                    \ifnum\exp=2
                        \node[anchor=east, xshift=1mm] at (\plotStartX + \figOff + \x * \doaDist - \tickLen,\plotStartY + \deltaY) {\footnotesize $10^{\text{-}\shiftExp}$};
                    \else
                \node[anchor=east, xshift=1mm] at (\plotStartX + \figOff + \x * \doaDist - \tickLen,\plotStartY + \deltaY) {\footnotesize $10^{\shiftExp}$};
                \fi
                \fi
            }{
                \def\tickLen{\minorTick}
            }
            \draw[line width=\tickWidth] (\plotStartX + \figOff + \x * \doaDist - \tickLen,\plotStartY + \deltaY) --  (\plotStartX + \figOff + \x * \doaDist, \plotStartY + \deltaY);
        }
    
    \else
    \pgfmathsetlengthmacro{\deltaY}{\figHeight / \numTicks}
    \foreach \y in {0, ..., \numTicks}{
        \pgfmathtruncatemacro\tmp{int(\y/2) * 2}
        \ifthenelse{\y = \numTicks}{
            \newcommand\metricTick{\high}
        }{
            \pgfmathsetmacro{\metricTick}{
        (\low + (\high - \low) / \numTicks * \y)
        }
        }
        \ifthenelse{\equal{\metric}{wer}}{
        \pgfmathsetmacro{\metricTick}{int(12.5 * \y)}
        \ifnum\tmp=\y
         \pgfmathsetmacro{\tickLen}{\majorTick}
            \node[anchor=east] at (\plotStartX + \figOff + \x * \doaDist - \tickLen,\plotStartY + \y * \deltaY) {\labelFont \text{\metricTick}};
        \else
            \pgfmathsetmacro{\tickLen}{\minorTick}
        \fi
        }{
            \ifnum\tmp=\y
            \pgfmathsetmacro{\tickLen}{\majorTick}
            \node[anchor=east] at (\plotStartX + \figOff + \x * \doaDist - \tickLen,\plotStartY + \y * \deltaY) {\labelFont \text{\metricTick}};
        \else
            \pgfmathsetmacro{\tickLen}{\minorTick}
        \fi
        }
        \draw[line width=\tickWidth] (\plotStartX + \figOff + \x * \doaDist - \tickLen,\plotStartY + \y * \deltaY) --  (\plotStartX + \figOff + \x * \doaDist,\plotStartY + \y * \deltaY);
    }
    \fi

    \ifnum\x=2
        \foreach \xTickOff / \xLabel in {
        0.05/{0}, 0.22/{1}, 0.4375/{2}, 0.66/{3}, 0.8775/{4}
        }{
            \draw[line width=\tickWidth] (\plotStartX + \figOff + \xTickOff * \figWidthTwo + \x * \doaDist,\plotStartY) --  (\plotStartX + \figOff + \xTickOff * \figWidthTwo + \x * \doaDist, \plotStartY - \majorTick);
            \node[anchor=north, rotate=0] at (\plotStartX + \figOff + \xTickOff * \figWidthTwo + \x * \doaDist,\plotStartY) {\labelFont \xLabel};
        }
    \else 
        \foreach \xLabel in {
        1,2,3,4
        }{
        \def\deltaSH{0.2}
        \pgfmathsetlengthmacro{\xTickOff}{((\xLabel-1) * 0.22 + 0.18)}
            \draw[line width=\tickWidth] (\plotStartX + \figOff + \xTickOff * \figWidthOne + \x * \doaDist,\plotStartY) --  (\plotStartX + \figOff + \xTickOff * \figWidthOne + \x * \doaDist, \plotStartY - \majorTick);
            \node[anchor=north, rotate=0] at (\plotStartX + \figOff + \xTickOff * \figWidthOne + \x * \doaDist,\plotStartY) {\labelFont \xLabel};
        }
    \fi
}

\node[xshift=2mm, yshift=-6mm] at (\plotStartX + \figWidth + \doaDist, \plotStartY) {
 spherical harmonics (SH) truncation order $N$ in \cref{eq:ambisonics_expansion}
};

\end{tikzpicture}
\vspace*{-20pt} %
\caption{
Compute (MACs) and enhancement performance (WER) with SpatialNet in real world recordings from an \texttt{Eigenmike64}.  
}
\label{fig:recordings}
\vspace*{-10pt}
\end{figure}

\textbf{\textit{Recorded dataset}}
\Cref{fig:recordings} displays the beamforming performance and complexity on our recorded dataset for varying ambisonics order $N$ with \mbox{SpatialNet} as mask estimator.
Note that all methods remain optimized for order $N=3$. 
While the low spatial resolution at $N=1$ is insufficient for enhancement in the multi-speaker meeting scenario, the weakly guided methods perform robustly across all higher orders (\picLegend{images/mvdr_beamformer/hatch_legend/fb_ambix.pdf},\,\picLegend{images/mvdr_beamformer/hatch_legend/ar_ambix.pdf},\,\picLegend{images/mvdr_beamformer/hatch_legend/dual_ambix.pdf}), demonstrating \textit{ambisonics-order-agnosticity}. 
Note that high ambisonics orders suffer from a limited bandwidth due to an ill-conditioned mode-strength inversion, yielding only negligible improvements from orders 3 to 4 at a significantly increased computational cost.
Even with the limited sample size, the combination of \ac{fb} and \ac{ar} input features consistently achieves superior enhancement~(\picLegend{images/mvdr_beamformer/hatch_legend/dual_ambix.pdf}). 
Listening tests, which are available on our project page\textsuperscript{\ref{project_page}}, reveal that autoregression helps SpatialNet to maintain \textit{speaker separability} in the latter part of the recordings.

\section{Conclusion}
\label{sec:conclusion}
In this work, we proposed a novel mask-based beamforming framework for moving speaker extraction in \acl{hoa} (\acs{hoa}) recordings.
By using a \acl{fb} (\acs{fb}) and the processed speech signal via temporal feedback as conditioning for neural mask estimation, we achieved robust enhancement solely based on the target speaker's starting direction.
In particular, we demonstrated how our \acl{ar} (\acs{ar}) incorporation of the processed speech improved the \textit{generalizability} to fast speaker movement and long-duration recordings. 
We could further show how the combination of \ac{fb} and \ac{ar} input features maintained superior enhancement across mask estimators and ambisonics orders, yielding an \textit{universally} applicable beamforming framework for arbitrary complexity constraints.
Real-world recordings complemented these findings under challenging acoustic conditions in an office meeting scenario.

\bibliographystyle{IEEEtran.bst}
\bibliography{strings,refs}

\end{document}